\documentclass[11pt,a4paper]{article}
\usepackage{amsmath,amssymb}
\usepackage{epsfig,graphicx}

\newtheorem {theorem}{Theorem}[section]
\newtheorem {lemma}[theorem]{Lemma}

\begin{document}
\title{\bf Quantization of Skyrmions}
\author{Steffen Krusch\footnote{{\bf e-mail}: 
S.Krusch@kent.ac.uk},
\\
\small{\em Institute of Mathematics, University of Kent,} \\
\small{\em Canterbury CT2 7NF, United Kingdom}
}
\date{October 2006}
\maketitle

\begin{abstract}
The Skyrme model is a nonlinear classical field theory which models
the strong interaction between atomic nuclei. In order to compare the
predictions of the Skyrme model with nuclear physics, it has to be
quantized. We show, summarizing earlier work, how the rational map
ansatz can be employed to calculate the Finkelstein-Rubinstein
constraints which arise during quantization. Then we give an overview
of current results on the quantum ground states in the Skyrme
model. We end with an outlook on future work.
\end{abstract}

\section{Introduction}

The Skyrme model is a classical field theory modelling the strong
interaction between atomic nuclei \cite{Skyrme:1961vq}. It has to be 
quantized in order to compare it to nuclear physics.
In \cite{Adkins:1983ya,Adkins:1984hy}, Adkins et al. quantized the
translational and rotational zero-modes of the $B=1$ Skyrmion for zero
and nonzero pion mass, respectively, and obtained good agreement with  
experiment.
A subtle point is that Skyrmions can be quantized as     
fermions as has been shown in \cite{Finkelstein:1968hy}. Solitons in
scalar field theories can consistently be quantized as fermions
provided that the fundamental group of configuration space has a
${\mathbb Z}_2$ subgroup generated by a loop in which two identical
solitons are exchanged.

The quantization of Skyrmions has a long history.
The $B=2$ Skyrmion with axial symmetry was quantized in
\cite{Braaten:1988cc, Kopeliovich:1988np, Verbaarschot:1987au} using
the zero-mode quantization. Later, the approximation was improved by
taking massive modes into account \cite{Leese:1995hb}.
The $B=3$ Skyrmion was first quantized in \cite{Carson:1991yv} and the
$B=4$ Skyrmion in \cite{Walhout:1992gr}. Irwin
performed a zero-mode quantization for $B=4-9$ \cite{Irwin:1998bs}    
using the monopole moduli space as an approximation for the Skyrmion  
moduli space. 
The physical predictions of the Skyrme model for various baryon numbers
were also discussed in \cite{Kopeliovich:2001yg}. 
Our aim here is to summarize the approach in
\cite{Krusch:2002by,Krusch:2005iq} and give an overview on current
progress and trends. Section \ref{Skyrme} gives a brief introduction to
the Skyrme model and describes the rational map ansatz from a more
topological point of view. Section \ref{Quantization} describes the
approach of Finkelstein and Rubinstein to the quantization of
Skyrmions, \cite{Finkelstein:1968hy}, and shows how the rational map
ansatz can be used to calculate the Finkelstein-Rubinstein
constraints. In Section \ref{Results}, we describe the calculation of
the quantum ground states in the Skyrme model using the zero-mode
approximation. We end with a discussion of which other physical 
effects have to be taken into account.

\section{The Skyrme Model}
\label{Skyrme}

The Skyrme model is a classical field theory of pions. The basic field
is the $SU(2)$ valued field\footnote{
Note that the Skyrme field can be written as 
$U({\bf x},t) = \exp \left(i \Pi_i({\bf x},t) \tau_i \right)$
where $\Pi_i({\bf x},t)$ are the pion fields and $\tau_i$ are the Pauli 
matrices. For small pion fields, the Skyrme Lagrangian can be expanded
in the $\Pi_i$-fields to give the standard Lagrangian for massive pions.}
$U({\bf x},t)$ where ${\bf x} \in {\mathbb
  R}^3$. The static solutions can be obtained by varying the
following energy
\begin{equation}
\label{energy}
E = \int \left( - \frac{1}{2} {\rm Tr} (R_i R_i)
- \frac{1}{16} {\rm Tr} ([R_i,R_j][R_i,R_j])
- m^2 {\rm Tr} (U-1) \right) d^3 x,   
\end{equation}
where $R_i = (\partial_i U) U^\dagger$ is a right invariant $su(2)$
valued current, and $m$ is a parameter proportional to the pion mass
$m_\pi$,
\cite{Battye:2004rw}. See also \cite{Kopeliovich:2005vg} for a
discussion of alternative pion mass terms. In (\ref{energy}) we
employed the ``geometric units'' in which length is measured in units
of $2/e f_\pi$ and energy in units of $f_\pi/4e$. The parameters
$f_\pi$ and $e$ are known as the pion decay constant and the Skyrme
constant, respectively.
In order to have finite energy, Skyrme fields
have to take a constant value, $U(|{\bf x}| = \infty) = 1$, at
infinity. Due to this boundary condition, all the points $|{\bf x}| =
  \infty$  can be identified and named ``$\infty$''. 
By a one-point
  compactification, the domain ${\mathbb R}^3$ together with
  ``$\infty$'' is topologically the three dimensional sphere
  $S^3$. Recall that the group $SU(2)$ as a manifold is also a 
three-sphere. 
Therefore, from a topological point of view the Skyrme field $U$ can
be regarded as a map $U: S^3 \to S^3$, and such maps are characterized
by an integer-valued winding number. 
This topological charge is interpreted as the baryon number,
which for our purposes can be thought of as the number of protons and
neutrons. It is given by the following integral
\begin{equation}
\label{B}
B= - \frac{1}{24 \pi^2} \int \epsilon_{ijk} {\rm Tr}(R_i R_j R_k) d^3 x.
\end{equation}
We will denote the configuration space of Skyrmions by $Q$. $Q$ splits
into connected components $Q_B$ labelled by the topological
charge. Furthermore, the energy of configurations in $Q_B$ is bounded
below by $E \ge 12 \pi^2 B,$ \cite{Faddeev:1976pg}.

\subsection{Rational Maps}

In this section, we describe the rational map ansatz \cite{Houghton:1998kg} 
which is a very successful approximation to minimal energy Skyrme 
configurations. The most convenient way for obtaining the explicit formula 
is the geometric approach of Manton \cite{Manton:1987xt, Krusch:2000gb}. 
In the following, we describe the construction in a more mathematical way, 
which will allow us to apply theorems from algebraic topology. The key 
idea is to view the rational map ansatz as a {\it suspension}. 

\begin{figure}
\begin{center}
\includegraphics[height=75mm,angle=0]{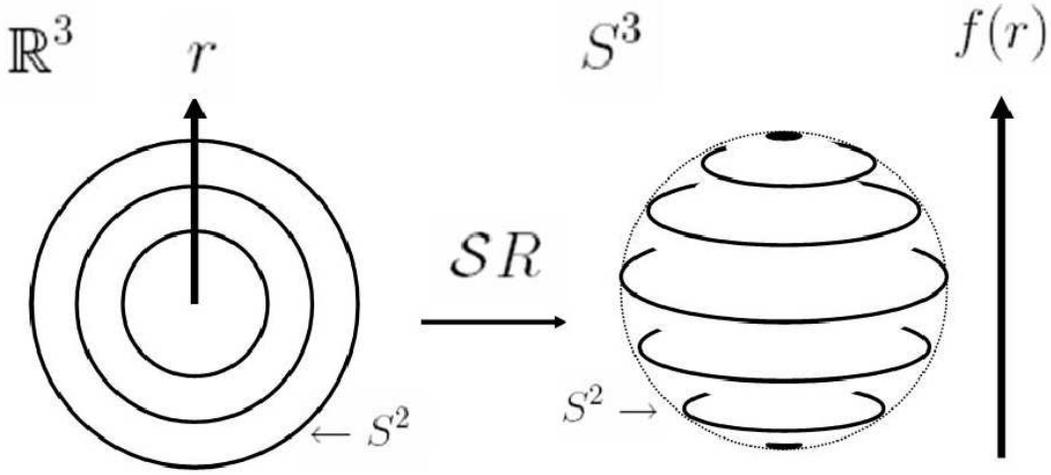}
\caption{An illustration of the suspension map for the rational map
  ansatz. Concentric spheres around the origin in ${\mathbb R}^3$ are
  mapped to spheres of latitude in $S^3$. Due to the boundary conditions, the
  sphere at $r=\infty$ can be regarded as one point. \label{fig1}}
\end{center}
\end{figure}

Given an interval $I = [0,1]$ and a manifold $M$, we can define the  
suspension ${\cal S} M$ by taking the Cartesian product $I \times M$ and 
then collapsing $\{0\} \times M$ to a point and also collapsing $\{1\} 
\times M$ to a point. A very important example is the suspension of 
spheres, namely, ${\cal S} S^n = S^{n+1}$. The standard polar coordinates 
are a smoothed-out version of a suspension. However, not only spaces, 
but also maps can be suspended. Consider the map 
\begin{equation}
R: S^2 \to S^2, z \mapsto R(z),
\end{equation}
which is a map between Riemann spheres. $z$ and $R$ are the standard 
complex coordinates obtained by stereographic projection. Now, both 
spheres can be suspended and we obtain $U = {\cal S} R$.
\begin{equation}
\label{suspension}
U: S^3 \to S^3, (r,z) \mapsto (f(r),R(z)).
\end{equation} 
The boundary conditions are $f(0) = \pi$ and $f(\infty) = 0$, so the end 
points of the ``interval'' $[0,\infty)$ are mapped to the endpoints of 
the interval $[0,\pi]$. For a graphical illustration of this
construction, see figure \ref{fig1}.

There is one further condition which makes the 
rational map ansatz particularly easy to use, namely, to consider only 
rational maps $R(z)$. Rational maps are holomorphic maps between Riemann 
spheres, and they can be written as ratios of two polynomials $p(z)$ and 
$q(z)$ which have no common factors, $R(z) = p(z)/q(z)$. The maximal 
polynomial degree of these polynomials $p(z)$ and $q(z)$ 
is also the topological degree of 
the rational map $R(z)$. Suspensions have good properties with respect 
to homotopy groups. Of particular importance is the Freudenthal 
suspension theorem \cite[Corollary 4.24]{Hatcher:2002}. 
One consequence of this 
theorem is that rational maps $R(z)$ of degree $B$ give rise to Skyrme 
fields $U$ also of degree $B$. 
When the ansatz (\ref{suspension}) is inserted into equation 
(\ref{energy}) we obtain
\begin{equation}
\label{Erational}
E = 4 \pi \int \left(r^2 {f^\prime}^2 + 2 B({f^\prime}^2+1) \sin^2 f 
+ {\cal I} \frac{\sin^4 f}{r^2} + 2 m^2 r^2(1-\cos f)
\right) {\rm d}r,
\end{equation}
where
\begin{equation}
\label{I}
{\cal I} = \frac{1}{4 \pi} \int \left( \frac{1+|z|^2}{1+|R|^2} 
\left|\frac{{\rm d}R}{{\rm d}z}\right| \right)^4 \frac{2 i~ {\rm d}z
{\rm d}{\bar z}}{(1+|z|^2)^2},
\end{equation}
and $B$ can be written as
\begin{equation}
\label{Brational}
B = \frac{1}{4 \pi} \int \left| \frac{{\rm d}R}{{\rm d}z} \right|^2
\frac{2i~{\rm d}z{\rm d}{\bar z}}{(1+|R|^2)^2}.
\end{equation}
Note that $B$ is the topological charge, and therefore an integer,
whereas ${\cal I}$ is a positive real number which depends on the given
rational map. From a geometric point of view, ${\cal I}$ measures the
angular strain orthogonal to the radial direction. It also has an
interesting interpretation as a possible Morse function on the moduli
space of monopoles \cite{Houghton:1998kg}.  
Now, we can first calculate the rational map which minimizes the integral 
${\cal I}$ and then solve the Euler-Lagrange equation for $f(r)$ subject 
to the boundary conditions $f(0) = \pi$ and $f(\infty) = 0$. 
The rational maps which minimize ${\cal I}$ for $m=0$ have been determined 
numerically in \cite{Battye:2001qn, Battye:2002wc} for all $B \le 40$. 

\begin{figure}
\begin{center}
\includegraphics[height=100mm,angle=0]{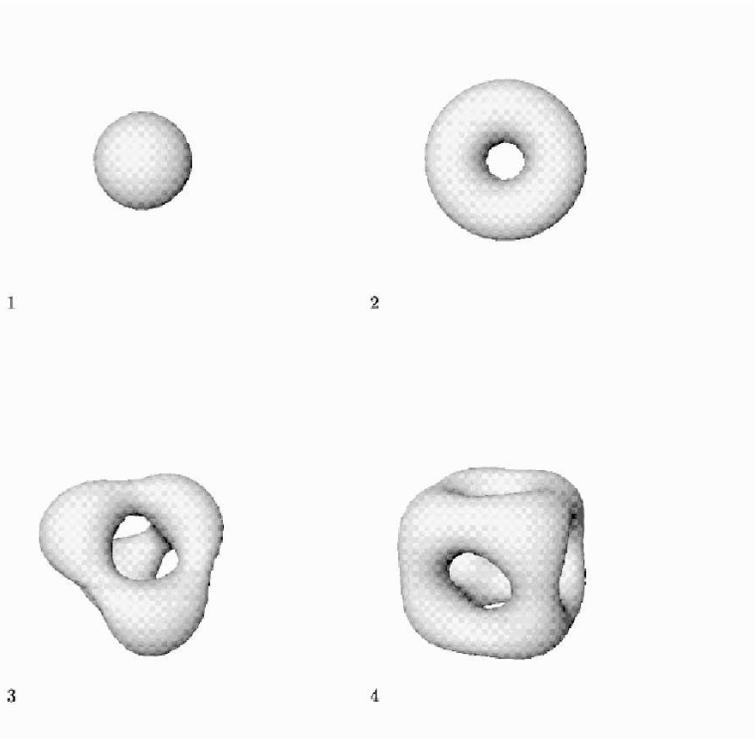}
\caption{Level sets of constant energy density for Skyrmions of baryon 
number $B=1, \dots 4.$ (Figure taken from \cite{Battye:1997nt}.) 
\label{fig2}}
\end{center}
\end{figure}

In figure \ref{fig2}, we show the lowest four ``Skyrmions'', that is 
the minimal energy configurations for baryon numbers $B=1, \dots, 4$. 
Here, we plot the level sets of constant energy density, namely the 
surfaces in ${\mathbb R}^3$ where the integrand of $(\ref{energy})$ has a 
given value. It is apparent that the configurations in figure 
\ref{fig2} are very ``symmetric''. The surface of constant energy 
density for $B=1$ is a sphere which has spherical symmetry, whereas $B=2$ 
has axial symmetry, and $B=3$ and $4$ have tetrahedral and cubic 
symmetry, respectively. In fact, not only the energy density is symmetric 
but also the Skyrme fields. 
By symmetry we mean that a rotation in space followed by a
rotation in target space leaves the Skyrmion invariant. Namely,   
\begin{equation}
\label{Usym}
U({\bf x} ) = A U \left(D(A^\prime) {\bf x} \right) A^\dagger,
\end{equation}
where $A$ and $A^\prime$ are $SU(2)$ matrices and $D(A^\prime)$ is the
associated $SO(3)$ rotation. These symmetries will play a very important 
role for the calculation of quantum ground states in the Skyrme model. 

Another curious fact about the pictures in figure \ref{fig2} are the 
``holes'', that is areas of particularly low energy density. The $B=4$
Skyrmion has six holes forming the faces of a cube, and the $B=3$
Skyrmion has $4$ holes. 
There is an easy way of understanding these ``holes''
from the rational map ansatz \cite{Houghton:1998kg}. 
The angular dependence of the energy
density in (\ref{Erational}) strongly depends on 
\begin{equation}
\label{Wronskian}
\frac{{\rm d} R}{{\rm d} z} = \frac{p^\prime(z) q(z) - q^\prime(z) 
p(z)}{q(z)^2}, 
\end{equation}
as can be seen from equations (\ref{I}) and (\ref{Brational}). The 
numerator of equation (\ref{Wronskian}) is known as the Wronskian and is 
generally a 
polynomial of degree $2B-2$. So, the zeros of the Wronskian give the 
faces of the Skyrme configurations. Similar polynomials also exist
whose zeros correspond to the edges and vertices of Skyrme
configurations.  

Note that the restriction that $R(z)$ is a
holomorphic map can be lifted and a generalized rational map ansatz
can be introduced \cite{Houghton:2001fe}. 
This generalized ansatz has been shown to improve the approximation 
for the energy significantly for the Skyrmions of degree $B=2,3,$ and
$4$ in figure \ref{fig2}, and it also captures the singularity
structure of Skyrmions better. 
However, it is difficult to use for higher baryon number, and from the 
point of view of discussing symmetries the original rational map ansatz 
is sufficient and easier to use.

\section{Quantization of Skyrmions}
\label{Quantization}

In the following, we recall the ideas of Finkelstein and Rubinstein
\cite{Finkelstein:1968hy} on how to quantize a scalar field theory and 
obtain fermions. We then show how to use the rational map ansatz to
calculate the Finkelstein-Rubinstein constraints.

\subsection{Finkelstein-Rubinstein Constraints}

In quantum field theory, there are two types of particles, namely {\em 
bosons} and {\em fermions}. If two identical particles are exchanged then 
nothing happens if these particles are {\em bosons}, 
whereas the wave function 
of the {\em fermions} changes by a factor of $(-1)$. 
When a {\em boson wave function} is rotated by 
$2 \pi$, it remains invariant. However, if a {\em fermion wave function} 
is rotated by $2 \pi$, then it changes by a factor of $(-1)$. The latter 
statement is a consequence of the spin-statistic theorem. In quantum field 
theory {\em bosons} are usually described by scalar, vector or tensor 
fields whereas {\em fermions} are represented by spinors.

\begin{figure}
\begin{center}
\includegraphics[height=85mm,angle=0]{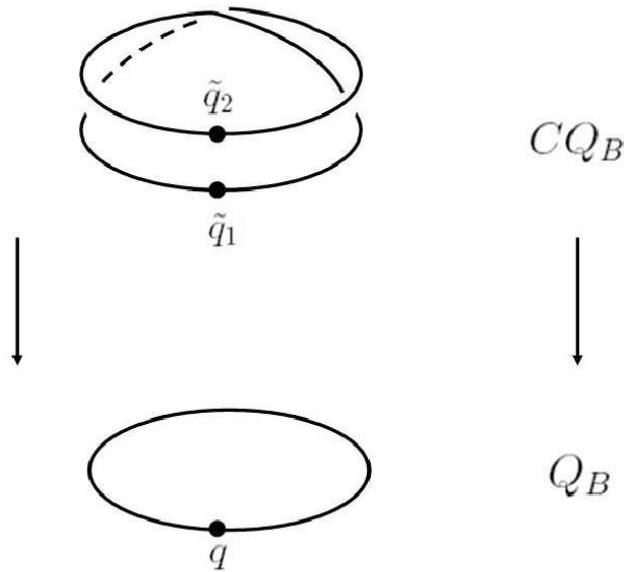}
\caption{This figure illustrates, schematically, the covering map. 
Both points ${\tilde
    q}_1$ and ${\tilde q}_2 \in CQ_B$ correspond to the same point $q
    \in Q_B$. Given a loop in configuration space starting at $q$ we
    can lift this loop to give a path in $CQ_B$ which will end at
    ${\tilde q}_2$, if the original loop is non-contractible, and
    will end at ${\tilde q}_1$ otherwise. \label{fig3}}
\end{center}
\end{figure}

So, how do we quantize the Skyrme model, which is a scalar field
theory, such that Skyrmions can model atomic nuclei, which consist of 
fermions? The key idea by Finkelstein and Rubinstein is to consider
wave functions on the covering space of configuration space. Note
that $Q_B$, the configuration space of Skyrme configurations of degree
$B$, has fundamental group $\pi_1(Q_B) \cong {\mathbb
  Z}_2$, in other words, its covering space $CQ_B$ is a double cover. 

Recall that one can think of the covering space $CQ_B$ as the space of 
equivalence classes of paths in the space $Q_B$. 
Figure \ref{fig3} illustrates this
construction. Consider the two points ${\tilde q}_1$ and ${\tilde q}_2
\in CQ_B$ which correspond to the point $q \in Q_B$. These two points
are joined by a path in $CQ_B$ which projects to a non-contractible
loop in $Q_B$.
This gives us a way of
examining continuous symmetries of Skyrme configurations. A
continuous symmetry of a configuration $q$ can be thought of as an
induced loop in configuration space. 
The symmetry also acts on covering space,
namely, if we apply the symmetry to ${\tilde q}_1$ we obtain ${\tilde
  q}_1$ if the induced loop is contractible, and ${\tilde q}_2$ if it
is not contractible.

We can now formally define a wave function $\psi$ as
\begin{equation}
\psi: CQ_B \to {\mathbb C}, {\tilde q} \mapsto \psi({\tilde q}),
\end{equation}
and impose the Finkelstein-Rubinstein constraint that 
$\psi({\tilde q}_1) = - \psi({\tilde q}_2)$ where ${\tilde q}_1$ and
${\tilde q}_2$ are defined as above.\footnote{
It is also consistent to impose the constraint 
$\psi({\tilde q}_1) = \psi({\tilde q}_2)$, 
but then all the quantum states are bosonic.} 
In the following, we are
particularly interested in loops which arise from the symmetries
(\ref{Usym}). Using the Finkelstein-Rubinstein constraint, we see that
a rotation by $\alpha$ around axis ${\bf n}$ in space followed by a
rotation by $\beta$ in target space around axis ${\bf N}$ gives 
\begin{equation}
\label{loop}
\exp\left(-i \alpha {\bf n} \cdot {\hat {\bf J}} \right)
\exp\left(-i \beta {\bf N} \cdot {\hat {\bf I}} \right)
\psi({\tilde q}) = \chi_{FR}~ \psi({\tilde q}),
\end{equation}
where
\begin{equation}
\chi_{FR} =
\left\{
\begin{array}{cc}
1 & {\rm if~the~induced~loop~is~contractible,} \\
-1& {\rm otherwise.}
\end{array}
\right.
\end{equation}
Here ${\hat {\bf J}}$ and ${\hat {\bf I}}$ are spin operators in space
and target space, respectively\footnote{For a discussion of the more
  subtle points about body-fixed and space-fixed angular momenta in
  this context see for example \cite{Braaten:1988cc,Krusch:2005bn}}.
For notational convenience, we will refer to a rotation in target
space as an {\em isorotation}. 
Before we discuss how to calculate the Finkelstein-Rubinstein phase
$\chi_{FR}$ using the rational map ansatz, 
we summarize some important and well-known results.
Giulini showed that a $2 \pi$ rotation of a Skyrmion gives rise to
$\chi_{FR} = (-1)$
if and only if the baryon number $B$ is odd, \cite{Giulini:1993gd}.
Finkelstein and Rubinstein showed in \cite{Finkelstein:1968hy} that a
$2 \pi$ rotation of a Skyrmion of degree $B$ is homotopic to an
exchange of two Skyrmions of degree $B$. This also implies that an
exchange of two identical Skyrmions gives rise to $\chi_{FR} = (-1)$
if and only if their baryon number $B$ is odd.
In \cite{Krusch:2002by} it was shown that a $2 \pi$ isorotation of a
Skyrmion also gives rise to $\chi_{FR} = (-1)$ if and only if the
baryon number $B$ is odd.
These results agree
with the physical intuition since atomic nuclei can be modelled by 
interacting point-like fermionic particles.

\subsection{Rational Maps and Finkelstein-Rubinstein constraints}

In this section, we show how to calculate the Finkelstein-Rubinstein
phase $\chi_{FR} \in \pi_1(Q_B)$ from the rational map ansatz
\cite{Krusch:2002by, Krusch:2005iq}. Using
the Freudenthal suspension theorem \cite[Corollary
  4.24]{Hatcher:2002}, the following theorem has been proved in
\cite{Krusch:2002by}. 
\begin{theorem}[S.K.] The rational map ansatz induces a surjective
  homomorphism \\ $\pi_1(Rat_B^*) \to \pi_1(Q_B^*).$
\end{theorem}
Here $Rat_B^*$ denotes the space of based rational maps of
degree $B$. Based rational maps satisfy the base point condition
$R(\infty) = 1$. 
The notation $Q_B^*$ just emphasizes that Skyrme configurations
are also based because $U(\infty) = 1$. The theorem gives us a way to
calculate $\chi_{FR}$ provided we know the fundamental group of
rational maps. The following theorem gives us the necessary information.
\begin{theorem}[Segal] $\pi_1(Rat_B^*) \cong {\mathbb Z}$ and it is
  generated by moving a zero once around a pole.
\end{theorem}
Note that a map $R \in Rat_B^*$ can be written as 
\begin{equation}
R(z) = \frac{z^B + a_{B-1} z^{B-1} + \dots + a_0}
{z^B + b_{B-1} z^{B-1} + \dots + b_0}
= \prod\limits_{i,j = 1}^{B} \frac{z-z_i}{z-p_j}.
\end{equation}
So, a based rational map $R(z)$ can be parameterized solely in terms of
its zeros $z_i$ and poles $p_j$. 
Given a loop $L$ that moves zeros $z_i$ and poles $p_j$ around in the
complex plane as a function of $\phi \in [0,\Phi]$, let
\begin{equation}
\label{Nint}
N(L) = \frac{i}{2 \pi} \sum\limits_{i,j=1}^B \int\limits_0^\Phi
\frac{\left(
z_i^\prime (\phi) - p_j^\prime (\phi)\right) {\rm d} \phi}
{\left(z_i(\phi) - p_j(\phi) 
\right)}.
\end{equation}
With this definition and using Cauchy's theorem it is straight forward
to prove the following lemma.
\begin{lemma} $N(L)$ is a homotopy invariant and counts the number of
  times zeros move around poles. Therefore, $N(L)$ provides an
  isomorphism $\pi_1(Rat_B^*) \to {\mathbb Z}.$
\end{lemma}
The above lemma gives us a way to
calculate $N(L)$ for any loop in the space of based rational maps
numerically. However, the fact that $N(L)$ is a homotopy invariant
allows us to make even more progress. Let us consider the axially
symmetric map  
\begin{equation}
\label{Raxial}
R(z) = \frac{z^B - b}{z^B+b}.
\end{equation}
We can now consider a loop as in (\ref{loop}), generated by a rotation
by an angle $\alpha$ around the $z$-axis followed by a rotation by an
angle $\beta$ around the
$X$-axis in target space. This choice of axes guarantees
that the whole loop respects the boundary condition 
$R(\infty) =  1$. 
For the rational map (\ref{Raxial}) it is easy to see how the zeros
and poles behave as the rotation angle goes from $0$ to $\alpha$ and
$\beta$, respectively. Using formula (\ref{Nint}) we can evaluate
$N(L)$ for this loop explicitly and obtain
\begin{equation}
\label{N}
N = \frac{B}{2 \pi} \left(B \alpha - \beta \right).
\end{equation}
Surprisingly, (\ref{N}) can be used in a much more general context.
\begin{theorem}[S.K.]\label{tN}
The value of $N$ for a given symmetry of a rational map $R \in Rat_B$
only depends on the rotation angle $\alpha$ and the isorotation angle
$\beta$, where the angles are defined such that $R(z_{-{\bf n}}) =
  R_{\bf N}$. It is given by (\ref{N}).
\end{theorem}
There is a choice for the sign of $\alpha$ which corresponds to the
choice of the rotation axis in space.\footnote{This sign choice can be
  fixed by imposing conditions on the signs of the components of the
  normal vector ${\bf n}$, see \cite{Krusch:2002by}.} Once the axis in
space is fixed the rational map determines the sign of the rotation
axis in target space via $R(z_{-{\bf n}}) = R_{\bf N}$. Here $z_{-{\bf
    n}}$ is the complex coordinate of the point $-{\bf n}$ and
similarly for $R_{{\bf N}}$. This condition is important since if the
wrong sign for $\beta$ is used in (\ref{N}) the value of 
$N$ might no longer be integer.
Now, we can express the Finkelstein-Rubinstein phase $\chi_{FR}$ in
terms of $N$ via the surjection
\begin{equation}
\chi_{FR} = (-1)^N.
\end{equation}

It is important to emphasize that theorem \ref{tN} can only be used for
Skyrme configurations which can be deformed into Skyrme fields
obtained from the rational map ansatz while keeping the relevant symmetry
\cite{Krusch:2005iq}.  
This is clearly the case for $2 \pi$ rotations. Hence, a $2 \pi$
rotation of a Skyrmion of degree $B$ gives 
\begin{equation}
\chi_{FR} = (-1)^{B^2},
\end{equation}
which is equal to $(-1)$ if and only if $B$ is odd. Similarly, a $2
\pi$ rotation in target space --- a $2 \pi$ isorotation --- gives rise to 
\begin{equation}
\chi_{FR} = (-1)^B,
\end{equation}
thus reproducing some of the results discussed at the end of the
previous subsection.

Often a Skyrme configuration $U$ can be approximated by assuming that
it consists of $K$ disjoint parts, each of which can be
approximated as a Skyrmion of degree $B_i$ centered around $X_i$. 
Then the truncated rational map ansatz can be written as
\begin{equation}
U({\bf x}) =
\left\{
\begin{array}{lc}
U_1({\bf x}), & {\rm for~}|{\bf x} - {\bf X}_1| < L_1, \\
\vdots & \vdots \\
U_K({\bf x}), & {\rm for~} |{\bf x} - {\bf X}_K| < L_K, \\
1,            & {\rm otherwise,}
\end{array}
\right.
\end{equation}
where the parameters $L_i$ are related to the size of the Skyrme
configurations and the individual parts $U_i$ are well approximated by
the rational map ansatz, subject to the boundary condition that
$U_i({\bf x}) = 1$ for $|{\bf x} - {\bf X}_i| = L_i$. It is clear from
formula (\ref{B}) that the baryon number of this configuration is given
by $B = B_1 + \dots + B_K$. 
We can now use this ansatz to calculate
the Finkelstein-Rubinstein constraints for Skyrme configuration which
are symmetric under a $C_n^k$ symmetry. By $C_n^k$ symmetry we mean a
rotation by $\alpha = \frac{2 \pi}{n}$ followed by a rotation by
$\beta = \frac{2 \pi k}{n}$ in target space. This $C_n^k$ symmetry
relates different Skyrmions $U_{B_i}$ with each other. For each
individual Skyrmion, there are two possibilities. Either the centre of
the Skyrmion lies on the symmetry axis, or the Skyrmion is part of an
$n$-gon of Skyrmions which transform into each other. Assume that
there are $l$ regular $n$-gons of Skyrmions with degree $B_i$ for
$i=1,\dots,l$ and $m$ Skyrmions of degree ${\tilde B}_j$ for
$j=1,\dots, m$ which are located on the symmetry axis. Then the
Finkelstein-Rubinstein constraints for this configuration are given by
\begin{equation}
\label{Ntr}
\chi_{FR} = (-1)^N,~~~{\rm where}~
N=\sum\limits_{i=1}^l B_i(n B_i - k) + 
\sum\limits_{j=1}^m {\tilde B}_j ({\tilde B}_j - k)/n.
\end{equation}

This approach is very well suited for calculating the
Finkelstein-Rubinstein constraints for (local) minima which are only
know numerically. Once, the minimal energy configuration has been
calculated, we have to determine its symmetry, and in particular
$C_n^k$ for a set of generators of the symmetry group. 
We confirm the symmetry
by starting with a symmetric configuration given, for example, by the
truncated rational map ansatz as initial condition and then letting
this configuration relax into the same final configuration. The
crucial point is that the relaxation method provides a homotopy from
the initial to the final configuration which is invariant under the
symmetry. Therefore, it is mathematically sound to calculate the
Finkelstein-Rubinstein constraints using formula (\ref{Ntr}).

\section{Results and Outlook}
\label{Results}

\begin{figure}
\begin{center}
\includegraphics[height=55mm,angle=0]{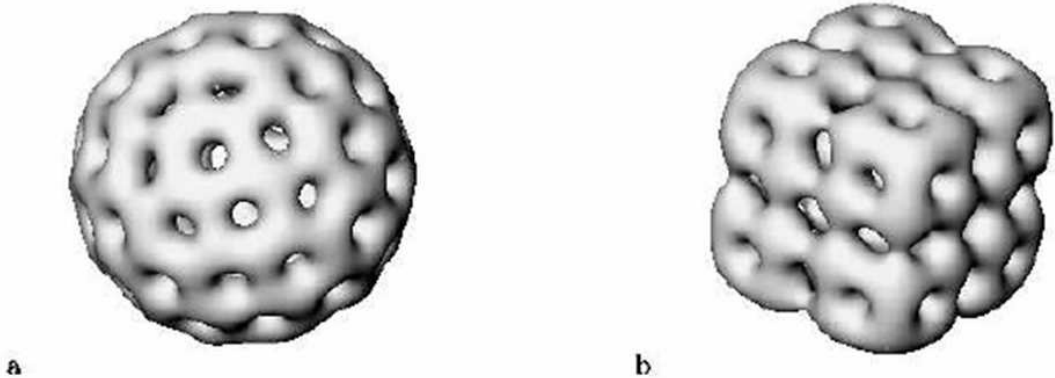}
\caption{This figure shows energy level sets for two $B=32$ Skyrme
  configurations which have
  very low energy. For pion mass $m=0$ the rational map configuration
  (a) has lower energy whereas for higher values of $m$ the ``cube'' (b)
  has lower energy. (Figure taken from \cite{Battye:2006tb}.) \label{fig4}}
\end{center}
\end{figure}

In this section, we present the results of the approach described in
the previous sections and also give a personal view on interesting
current and future developments.\footnote{
Here we focus only on the $SU(2)$ Skyrme model and do not consider the
interesting effects of gravity. Note, however, that the Skyrme model
can be consistently quantized fermionically  when the domain is any
compact, orientable $3$-manifold \cite{Auckly:2004yg}.}

The simplest nontrivial application of this approach is to quantize
the zero-modes around the classical minimal energy configurations in
each sector $Q_B$, see \cite{Irwin:1998bs, Krusch:2002by}. 
Then, the quantum ground state
$|J\rangle | I \rangle$ is given by the lowest values of the angular
momentum quantum numbers $J$ and $I$ which are compatible with the
Finkelstein-Rubinstein constraints for the symmetries of the classical
minimal energy configuration taken from \cite{Battye:2001qn}. 
For $B=1, \dots, 4$ the results are
promising. The Skyrme model reproduces the correct ground state\footnote{
Recall that the Skyrme model only captures the strong interaction, so
that proton and neutron are degenerate in energy in the Skyrme model.
By gauging the Skyrme model, the electromagnetic interaction can be
taken into account, see e.g. \cite{Piette:1997ny}.}
for $^1_1$H ($|\frac{1}{2}\rangle |\frac{1}{2}
\rangle$), the deuteron $^2_1$H ($|1\rangle |0\rangle$), $^3_2$He
($|\frac{1}{2}\rangle | \frac{1}{2} \rangle$),  
and the alpha-particle $^4_2$He ($|0\rangle|0\rangle$). The ground
states corresponding to $B= 8, 12, 16, 20$ which are particularly
stable are also correctly predicted as $|0\rangle|0\rangle$
states. In \cite{Krusch:2005iq} states with even nucleon
  number $B$ have been
  examined further. The rational map ansatz gives restriction on what
  $C_n^k$ symmetry can be realized. These restrictions are compatible
  with the phenomenological even-even and odd-odd nucleons rule.
However, for $B=5$ the predicted
ground state does not agree with experiment. One might argue that this
is not such a problem, since there is no stable nucleus with $5$
nuclei, however, this result marks the beginning of a trend. The
predictions for odd nuclei are not reliable at all. In order to
understand what is going wrong, we have to discuss the approximations
in our approach in more detail. The zero-mode approximation neglects
any deformations and vibrations of the Skyrme field and assumes that
the Skyrmion rotates like a rigid body. Schematically, we obtain the
following formula for the energy of a given state.
\begin{equation}
\label{EIJ}
E \approx M_{{\rm classical}} + \frac{\hbar^2}{2 \Theta_J} J(J+1)
+ \frac{\hbar^2}{2 \Theta_I} I(I+1).
\end{equation}
The above formula helps us to understand why the states with vanishing
spin $J$ and vanishing isospin $I$ are modelled quite successfully by the 
classical minimals, namely, the contributions of the second and the
third term vanish. For nucleon numbers $B$ which are divisible by two
but not by four, the predictions only fail for $B=10, 18,$ and $22$.

For fermions, however, the lowest possible quantum numbers are
$J=\frac{1}{2}$, $I = \frac{1}{2}$, so that all three terms contribute
in (\ref{EIJ}). These further contributions to the energy make it
necessary to take local minima into account, when calculating quantum
ground states. It is also important to allow the Skyrmions to deform
while they are spinning, see \cite{Battye:2005nx} and references
therein. A related approach constructing a quantum hamiltonian for
spinning Skyrmions is described in
\cite{Houghton:2005iu}. Much work still needs to be done to
understand spinning Skyrmions, both classically and quantum
mechanically. In the former case, progress has been made by
applying the theory of relative equilibria (work in progress).

However, there is another important effect, namely the effect of the
pion mass term which is the last term in (\ref{energy}). In
\cite{Battye:2004rw}, it was shown that the physical value of the pion
mass leads to shell-like configurations becoming unstable to
squashing. While the minimals for small $B \le 8$ hardly change, many
new (local) minimal energy configurations have been found for larger $B$, see
\cite{Battye:2006tb, Battye:2006na, Houghton:2006ti}. These new
configurations are no longer shell-like, but often seem to be composed
of $B=4$ cubes. The relation to the phenomenological alpha particle
model is discussed in \cite{Battye:2006na}. There is hope that the
Skyrme model will one day be able to describe the low energy excited
states for example for $B=12$ or $B=16$ as rotational bands related to
different local minima in the same way as in the 
alpha particle model. In order to
achieve this aim, we need a better understanding of the classical
solutions, possibly including saddle point solutions in the Skyrme
model \cite{Krusch:2004uf}. 
This can be achieved by numerical simulations in connection
with various analytic approximations, such as the instanton ansatz
\cite{Atiyah:1989dq}, including the version in hyperbolic space which takes
account of the pion mass term \cite{Atiyah:2004nh}, and generalizations of
the rational map ansatz. 

We also need to address the important problem of
fixing the values of the Skyrme parameters $f_\pi$, $e$ and $m_\pi$. 
The original, and most widely used, set of Skyrme
parameters has been proposed in \cite{Adkins:1983ya,Adkins:1984hy},
by matching to the proton and the Delta mass. However, it has been
shown that this matching condition can be considered to be an artifact
of the rigid rotator approximation, \cite{Battye:2005nx}. The studies
in \cite{Battye:2006na} suggest that the effective pion mass $m$
should have the rather large value of $m \approx 1$. In
\cite{Kopeliovich:2004pd}, a 30\% lower value of the Skyrme
parameter was suggested in order to match a large range of nuclei masses to
experimental data. A similar conclusion was reached in
\cite{Manton:2006tq} by considering the electromagnetic properties of the
quantized $B=6$ Skyrmion, describing $_3^6$Li.

Although there are many problems to overcome, there is cautious
optimism that the Skyrme model can really teach us something about the
behaviour of small to medium sized nuclei.

\section*{Acknowledgments}

S. K. would like to thank the organizers of Quarks-2006 for a very
interesting conference. The author is grateful to N. S. Manton, J. M. Speight
and S. W. Wood for fruitful discussions and acknowledges the
hospitality of DAMTP, Cambridge. 
S. K. also wants to thank C.-Y. K. Kuan for help with the figures.

\renewcommand{\baselinestretch}{1}
\addcontentsline{toc}{section}{Bibliography}

\begin{thebibliography}{10}

\bibitem{Skyrme:1961vq}
T.~H.~R. Skyrme, {\em A {N}onlinear field theory\/}, Proc. Roy. Soc. Lond. {\bf
  A260}: 127 ({\bf 1961}),

\bibitem{Adkins:1983ya}
G.~S. Adkins, C.~R. Nappi and E.~Witten, {\em Static properties of nucleons in
  the {S}kyrme model\/}, Nucl. Phys. {\bf B228}: 552 ({\bf 1983}),

\bibitem{Adkins:1984hy}
G.~S. Adkins and C.~R. Nappi, {\em The {S}kyrme model with pion masses\/},
  Nucl. Phys. {\bf B233}: 109 ({\bf 1984}),

\bibitem{Finkelstein:1968hy}
D.~Finkelstein and J.~Rubinstein, {\em Connection between spin, statistics, and
  kinks\/}, J. Math. Phys. {\bf 9}: 1762 ({\bf 1968}),

\bibitem{Braaten:1988cc}
E.~Braaten and L.~Carson, {\em The deuteron as a toroidal {S}kyrmion\/}, Phys.
  Rev. {\bf D38}: 3525 ({\bf 1988}),

\bibitem{Kopeliovich:1988np}
V.~B. Kopeliovich, {\em Quantization of the rotations of axially symmetric
  systems in the {S}kyrme model\/}, Sov. J. Nucl. Phys. {\bf 47}: 949--953
  ({\bf 1988}),

\bibitem{Verbaarschot:1987au}
J.~J.~M. Verbaarschot, {\em Axial symmetry of bound baryon number two solution
  of the {S}kyrme model\/}, Phys. Lett. {\bf B195}: 235 ({\bf 1987}),

\bibitem{Leese:1995hb}
R.~A. Leese, N.~S. Manton and B.~J. Schroers, {\em Attractive channel
  {S}kyrmions and the deuteron\/}, Nucl. Phys. {\bf B442}: 228 ({\bf 1995}),
  {\ttfamily{<hep-ph/9502405>}},

\bibitem{Carson:1991yv}
L.~Carson, {\em B = 3 nuclei as quantized multiskyrmions\/}, Phys. Rev. Lett.
  {\bf 66}: 1406 ({\bf 1991}),

\bibitem{Walhout:1992gr}
T.~S. Walhout, {\em Quantizing the four baryon skyrmion\/}, Nucl. Phys. {\bf
  A547}: 423 ({\bf 1992}),

\bibitem{Irwin:1998bs}
P.~Irwin, {\em Zero mode quantization of multi-{S}kyrmions\/}, Phys. Rev. {\bf
  D61}: 114024 ({\bf 2000}), {\ttfamily{<hep-th/9804142>}},

\bibitem{Kopeliovich:2001yg}
V.~B. Kopeliovich, {\em Characteristic predictions of topological soliton
  models\/}, J. Exp. Theor. Phys. {\bf 93}: 435--448 ({\bf 2001}),
  {\ttfamily{<hep-ph/0103336>}},

\bibitem{Krusch:2002by}
S.~Krusch, {\em {Homotopy of rational maps and the quantization of
  Skyrmions}\/}, Ann. Phys. {\bf 304}: 103--127 ({\bf 2003}),
  {\ttfamily{<hep-th/0210310>}},

\bibitem{Krusch:2005iq}
S.~Krusch, {\em {Finkelstein-Rubinstein constraints for the Skyrme model with
  pion masses}\/}, Proc. Roy. Soc. {\bf A462}: 2001--2016 ({\bf 2006}),
  {\ttfamily{<hep-th/0509094>}},

\bibitem{Battye:2004rw}
R.~Battye and P.~Sutcliffe, {\em Skyrmions and the pion mass\/}, Nucl. Phys.
  {\bf B705}: 384--400 ({\bf 2005}), {\ttfamily{<hep-ph/0410157>}},

\bibitem{Kopeliovich:2005vg}
V.~B. Kopeliovich, B.~Piette and W.~J. Zakrzewski, {\em {Mass terms in the
  Skyrme model}\/}, Phys. Rev. {\bf D73}: 014006 ({\bf 2006}),
  {\ttfamily{<hep-th/0503127>}},

\bibitem{Faddeev:1976pg}
L.~D. Faddeev, {\em Some comments on the many dimensional solitons\/}, Lett.
  Math. Phys. {\bf 1}: 289 ({\bf 1976}),

\bibitem{Houghton:1998kg}
C.~J. Houghton, N.~S. Manton and P.~M. Sutcliffe, {\em Rational maps, monopoles
  and {S}kyrmions\/}, Nucl. Phys. {\bf B510}: 507 ({\bf 1998}),
  {\ttfamily{<hep-th/9705151>}},

\bibitem{Manton:1987xt}
N.~S. Manton, {\em Geometry of {S}kyrmions\/}, Commun. Math. Phys. {\bf 111}:
  469 ({\bf 1987}),

\bibitem{Krusch:2000gb}
S.~Krusch, {\em {${S}^3$ Skyrmions and the Rational Map Ansatz}\/},
  Nonlinearity {\bf 13}: 2163 ({\bf 2000}), {\ttfamily{<hep-th/0006147>}},

\bibitem{Hatcher:2002}
A.~Hatcher, {\em Algebraic Topology\/}, Cambridge University Press, Cambridge
 ({\bf 2002}).

\bibitem{Battye:2001qn}
R.~A. Battye and P.~M. Sutcliffe, {\em Skyrmions, fullerenes and rational
  maps\/}, Rev. Math. Phys. {\bf 14}: 29 ({\bf 2002}),
  {\ttfamily{<hep-th/0103026>}},

\bibitem{Battye:2002wc}
R.~A. Battye, C.~J. Houghton and P.~M. Sutcliffe, {\em {Icosahedral
  Skyrmions}\/}, J. Math. Phys. {\bf 44}: 3543--3554 ({\bf 2003}),
  {\ttfamily{<hep-th/0210147>}},

\bibitem{Battye:1997nt}
R.~A. Battye and P.~M. Sutcliffe, {\em Multi-soliton dynamics in the {S}kyrme
  model\/}, Phys. Lett. {\bf B391}: 150 ({\bf 1997}),
  {\ttfamily{<hep-th/9610113>}},

\bibitem{Houghton:2001fe}
C.~J. Houghton and S.~Krusch, {\em Folding in the {S}kyrme model\/}, J. Math.
  Phys. {\bf 42}: 4079 ({\bf 2001}), {\ttfamily{<hep-th/0104222>}},

\bibitem{Krusch:2005bn}
S.~Krusch and J.~M. Speight, {\em {Fermionic quantization of Hopf solitons}\/},
  Commun. Math. Phys. {\bf 264}: 391--410 ({\bf 2006}),
  {\ttfamily{<hep-th/0503067>}},

\bibitem{Giulini:1993gd}
D.~Giulini, {\em On the possibility of spinorial quantization in the {S}kyrme
  model\/}, Mod. Phys. Lett. {\bf A8}: 1917 ({\bf 1993}),
  {\ttfamily{<hep-th/9301101>}},

\bibitem{Battye:2006tb}
R.~Battye and P.~Sutcliffe, {\em Skyrmions with massive pions\/}, Phys. Rev.
  {\bf C73}: 055205 ({\bf 2006}), {\ttfamily{<hep-th/0602220>}},

\bibitem{Auckly:2004yg}
D.~Auckly and J.~M. Speight, {\em {Fermionic quantization and configuration
  spaces for the Skyrme and Faddeev-Hopf models}\/}, Commun. Math. Phys. {\bf
  263}: 173--216 ({\bf 2006}), {\ttfamily{<hep-th/0411010>}},

\bibitem{Piette:1997ny}
B.~M. A.~G. Piette and D.~H. Tchrakian, {\em {Topologically stable soliton in
  the U(1) gauged Skyrme model}\/}, Phys. Rev. {\bf D62}: 025020 ({\bf 2000}),
  {\ttfamily{<hep-th/9709189>}},

\bibitem{Battye:2005nx}
R.~A. Battye, S.~Krusch and P.~M. Sutcliffe, {\em {Spinning Skyrmions and the
  Skyrme parameters}\/}, Phys. Lett. {\bf B626}: 120--126 ({\bf 2005}),
  {\ttfamily{<hep-th/0507279>}},

\bibitem{Houghton:2005iu}
C.~Houghton and S.~Magee, {\em A zero-mode quantization of the {S}kyrmion\/},
  Phys. Lett. {\bf B632}: 593--596 ({\bf 2006}), {\ttfamily{<hep-th/050909>}},

\bibitem{Battye:2006na}
R.~Battye, N.~S. Manton and P.~Sutcliffe, {\em Skyrmions and the alpha-particle
  model of nuclei\/}  ({\bf 2006}), {\ttfamily{<hep-th/0605284>}},

\bibitem{Houghton:2006ti}
C.~Houghton and S.~Magee, {\em The effect of pion mass on {S}kyrme
  configurations\/}  ({\bf 2006}), {\ttfamily{<hep-th/0602227>}},

\bibitem{Krusch:2004uf}
S.~Krusch and P.~Sutcliffe, {\em {Sphalerons in the Skyrme model}\/}, J. Phys.
  {\bf A37}: 9037 ({\bf 2004}), {\ttfamily{<hep-th/0407002>}},

\bibitem{Atiyah:1989dq}
M.~F. Atiyah and N.~S. Manton, {\em Skyrmions from instantons\/}, Phys. Lett.
  {\bf B222}: 438 ({\bf 1989}),

\bibitem{Atiyah:2004nh}
M.~Atiyah and P.~Sutcliffe, {\em Skyrmions, instantons, mass and curvature\/},
  Phys. Lett. {\bf B605}: 106--114 ({\bf 2005}), {\ttfamily{<hep-th/0411052>}},

\bibitem{Kopeliovich:2004pd}
V.~B. Kopeliovich, A.~M. Shunderuk and G.~K. Matushko, {\em Mass splittings of
  nuclear isotopes in chiral soliton approach\/}, Phys. Atom. Nucl. {\bf 69}:
  120--132 ({\bf 2006}), {\ttfamily{<nucl-th/0404020>}},

\bibitem{Manton:2006tq}
N.~S. Manton and S.~W. Wood, {\em {Reparametrising the Skyrme Model using the
  Lithium-6 Nucleus}\/}  ({\bf 2006}), {\ttfamily{<hep-th/0609185>}}.

\end{thebibliography}

\begin{small}

\end{small}
\label{lastref}

\end{document}